# Enhancing Food Safety in Supply Chains: The Potential Role of Large Language Models in Preventing Campylobacter Contamination

*Asaf Tzachor* [1,2] *

Author information

Affiliations

[1] Reichman University, School of Sustainability, Herzliya, Israel

Corresponding author *

Asaf Tzachor, atzachor@runi.ac.il

## Abstract

Foodborne diseases pose a significant global public health challenge, primarily driven by bacterial infections. Among these, Campylobacter spp. is notable, causing over 95 million cases annually. In response, the Hazard Analysis and Critical Control Points (HACCP) system, a food safety management framework, has been developed and is considered the most effective approach for systematically managing foodborne safety risks, including the prevention of bacterial contaminations, throughout the supply chain. Despite its efficacy, the adoption of HACCP is often incomplete across different sectors of the food industry. This limited implementation can be attributed to factors such as a lack of awareness, complex guidelines, confusing terminology, and insufficient training on the HACCP system's implementation. This study explores the potential of large language models (LLMs), specifically generative pre-trained transformers (GPTs), to mitigate Campylobacter contamination across four typical stages of the supply chain: primary production, food processing, distribution and retail, and preparation and consumption. While the interaction between LLMs and food safety presents a promising potential, it remains largely underexplored. To demonstrate the possible applications of LLMs in this domain, we further configure an open-access customized GPT trained on the FAO's HACCP toolbox and the 12 steps of HACCP implementation, and test it in the context of commercial food preparation. The study also considers critical barriers to implementing GPTs at each step of the supply chain and proposes initial measures to overcome these obstacles.

## Graphical abstract

Fifteen (15) potential applications of Large Language Models (LLMs) to reduce Campylobacter contamination along four typical steps of the food supply chain

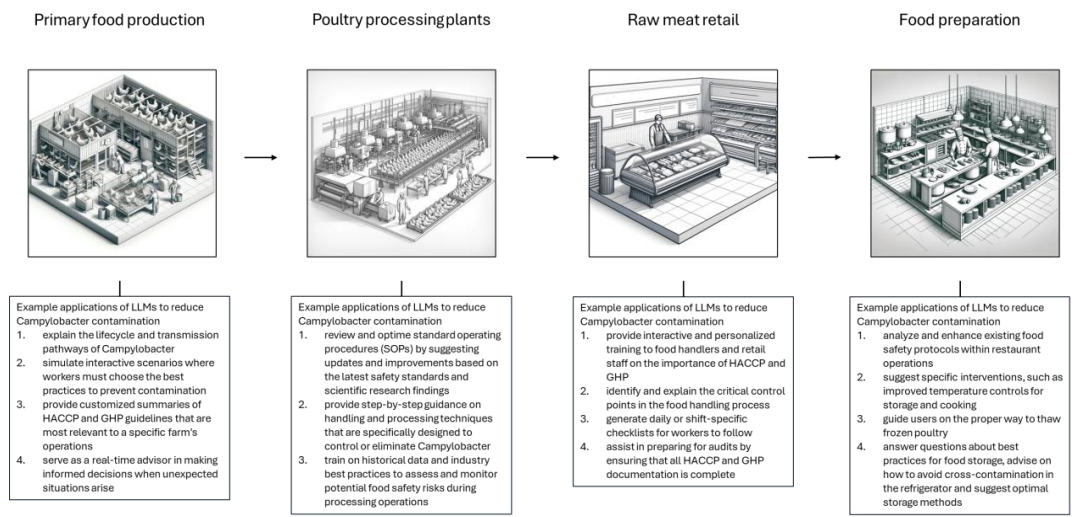

# 1. Introduction

Food safety is a global public health concern, and a primary pillar of food security. Formally, food safety is defined as both a scientific discipline and as a set of practices seeking to ensure food is free from pathogens (e.g., bacteria, viruses and parasites) and other contaminants (e.g., pesticide residues), to prevent foodborne illness, and to warrant it fit for human consumption [1,2,3,4]. As such, food safety and hygiene span the entire food supply chain, from "farm to fork", including primary production (of both plant-based and animal-based foods), food processing, distribution and retail, and handling and preparation [5]. From the perspective of food security, if food items are not safe for consumption, they cannot fulfill their nutritional purpose and contribute to food security, even if these items are available, affordable, and accessible. As the Global Hunger and Food Security Initiative asserts, "if it isn't safe, it isn't food." [6]

Previous studies have shown that the prevalence of foodborne illnesses increased over recent years, adversely affecting the lives and livelihoods of populations across developing and developed regions [7,8,9,10,11]. Landmark reports by the World Health Organization on the global incidence of foodborne diseases determined that unsafe food consumption leads to 600 million instances of foodborne diseases and 420,000 deaths each year. Additionally, out of 56 million total annual deaths, 7.5% (420,000) are attributed to foodborne illnesses [12,13]. Data reveals that 7.69% of the global population (7.8 billion) suffer from foodborne diseases annually. Comparatively, this is almost a third of the annual deaths caused by road traffic accidents worldwide, which total 1.35 million [14].

Globally, bacterial infections are the most common cause of foodborne illnesses, with 226,526,634 cases reported, followed by viral (138,513,782 cases) and parasitic infections (10,284,561 cases) [15]. Among these, norovirus stands out with the highest annual cases at 124,803,946 and 34,929 deaths, yielding a fatality rate of 0.028. Norovirus outbreaks have been linked to vegetables, fruits, raw oysters, and processed meats, emphasizing the need to monitor it as a significant risk in meat products [16]. Hepatitis A virus also presents a significant risk, with 13,709,836 cases and a fatality rate of 0.202, highlighting viral dangers associated with meat consumption.

Among bacteria, Campylobacter leads with 95,613,970 cases annually, although its fatality rate is low at 0.022, followed by Salmonella with 78,707,591 cases. In contrast, L. monocytogenes, although less frequent, poses a high risk with a fatality rate of 22.41%, indicating its severe impact relative to other bacteria [17]. Both Campylobacter and L. monocytogenes are primarily associated with meat and processed animal products [18]. Regarding parasites, Toxoplasma gondii still accounts for a high number of foodborne illnesses, with 10,280,089 cases and 684 deaths annually. It primarily affects animals but can infect humans through the consumption of uncooked or raw meat, underscoring the persistent risk of parasitic infections [19].

While the economic toll of foodborne diseases is difficult to calculate accurately, the WHO has estimated that 31 foodborne hazards considered priority posed a burden of about 33 million disability adjusted life years (DALYs) in 2010 [20]. Furthermore, climate change and its resulting alternations in abiotic factors (e.g., temperature and humidity, precipitation patterns) is expected to increase incidents of bacterial contamination and, ultimately, the prevalence of foodborne diseases, thereby posing further challenges in certain regions [21].

To mitigate food safety risks, and foodborne diseases, the Hazard Analysis and Critical Control Points (HACCP), a Food Safety Management System (FSMS), has been introduced in the 1960s and since 1993, the application of the HACCP system has been recommended by the Codex Alimentarius published by the Food and Agriculture Organization (FAO) of the United Nations [22].

The HACCP system is widely recognized by the international food safety community as the best method for systematically identifying, assessing, and controlling foodborne safety risks along the supply chain [23, 24,25]. A holistic approach to ensure food safety, sanitation, and hygiene across the supply chain, the HACCP systems also promote the Good Hygiene Practices (GHP). Although the concept of Good Hygiene Practices (GHP) predates the development of HACCP, the two are closely related and considered complementary food safety management approaches (see Box 1.).

---

Box 1. GHP, HACCP principles, and the 12-step HACCP FSMS

Good Hygiene Practices (GHP) are a set of conditions and activities, delineated in the Codex Alimentarius General Principles of Food Hygiene (CXC 1-1969), that are necessary to maintain a hygienic environment throughout the food supply chain. They serve as the foundation for effective FSMS and HACCP. GHP covers a wide range of practices, including proper personal hygiene, cleaning and sanitizing procedures, pest control, waste management, and the appropriate design and maintenance of equipment and facilities.

Hazard Analysis and Critical Control Points (HACCP) is a systematic preventive approach to food safety that identifies physical, chemical, and biological hazards in production processes that can cause the finished product to be unsafe. It designs measurements to reduce these risks to a safe level. HACCP is a framework for monitoring the critical control points identified as risks to food safety. Implementing HACCP helps prevent, eliminate, or reduce these hazards to an acceptable level.

The 12 steps involved in implementing an HACCP FSMS include: (1) Assemble the HACCP team with appropriate product-specific knowledge and expertise, (2) Describe the food product in detail, including ingredients, packaging, and brand names, (3) Identify the intended use and consumers of the food product, (4) Construct a flow diagram of the production process from receiving to shipping, (5) Verify the accuracy of the flow diagram on-site, (6) Conduct a hazard analysis by identifying potential biological, chemical, and physical hazards associated with each process step, (7) Determine the critical control points (CCPs) in the process, (8) Establish critical limits for preventive measures at each CCP, (9) Establish monitoring procedures for CCPs, (10) Establish corrective actions when monitoring shows a CCP is out of control, (11) Establish verification procedures to ensure the HACCP system is working correctly, and (12) Establish effective record-keeping procedures documenting the HACCP system.

Despite its efficacy, the adoption of HACCP is often incomplete [26,27,28,29,30]. The limited implementation of the framework is oftentimes attributed to lack of knowledge and understanding of these frameworks among institutions along the food chain [31,32,33,34,35], with a few studies further citing insufficiency of simple guidelines, confusion in terminology, lack of training in how to implement such frameworks as additional implementation barriers, although to varying degrees [36]. Indeed, previous research has long substantiated that HACCP principles can be realized only if the people responsible for implementation have the knowledge and expertise to apply it effectively [37], [38,39].

Against this backdrop, in this paper we explore the potential of large language models (LLMs), namely generative pre-trained transformers (GPT), to simplify scientific knowledge and practices pertaining to HACCP FSMS across a typical food supply chain.

In line with previous literature, and prevalence of bacteria-related foodborne illnesses, in this exploratory study we examine the potential of LLMs by focusing on *Campylobacter spp*. Campylobacter is chosen as the focus of this study for several reasons. Among bacterial foodborne illness sources, Campylobacter leads with over 95 million cases annually. The bacteria are known to cause enteritis and gastroenteritis in both adults and children. Recently, cases of Campylobacteriosis in high-income countries have surpassed those caused by traditional enteric bacteria. The bacteria are found in the digestive tract of infected patients about 3-4 times more often than Salmonella or Escherichia coli. In low- and middle-income countries, there's evidence suggesting an increase in Campylobacter infections [40,41,42].

Contamination can occur at various stages of the food supply chain including during animal rearing, particularly in poultry, slaughtering, processing, retailing, and domestic food preparation, with varying occurrence rates contingent on the type of meat and handling practices [43,44] The sporadic nature of the infection and cross-contamination make it challenging to trace the source of exposure. Consequently, many countries have implemented preventive measures over the past decade. Scientific advancements have improved our understanding of Campylobacter infections, including their causes and the role of immunity. Despite these efforts, Campylobacteriosis remains difficult to prevent, and infection rates remain high worldwide [45].

The dependency on the poultry industry for providing high-quality easily digestible protein, essential vitamins such as phosphorus, B-complex vitamins, and iron, zinc, and selenium, to billions worldwide [46,47,48], exacerbates the predicament further.

Focusing on a single pathogen, this study does not examine other biological hazards (e.g., Salmonella) pesticide residues, food additives and preservatives (chemical hazards), foreign objects such as plastic contaminants (physical hazards), nor allergens.

While LLMs and GPTs, emerging and seemingly potent artificial intelligence (AI) technologies, present an intriguing opportunity to enhance food safety, the 'LLM – food safety' interface has been largely unexplored.

In addition, to demonstrate the potential applications of LLMs in this domain, we create an open-access customized GPT (using OpenAI's My GPTs, ChatGPT configuration platform), trained on the FAO's GHP and HACCP Toolbox for Food Safety [49], as well as the 12 steps involved in implementing an HACCP FSMS (described in Box 1), to provide comprehensible, user-friendly, and contaminant specific (Campylobacter), actionable recommendations for restaurant kitchen employees. Box 2 provides the configuration details of the chatbot.

---

Box 2. Configuration details for Campylobacter Prevention in Restaurant Kitchens

Campylobacter Prevention in Restaurant Kitchens is a publicly available chatbot, configured with OpenAI's My GPT interface, at: https://chatgpt.com/g/g-LI7Y7dhko-campylobacter-prevention-in-restaurant-kitchens.

The GPT is configured as follows:

Description: A user-friendly GPT-advisor helping restaurant kitchen employees implement the Hazard Analysis and Critical Control Points (HACCP) system and the Good Hygiene Practices (GHP) to avoid Campylobacter contamination in poultry during food preparation.

Instructions: This GPT is designed to advise restaurant kitchen employees who handle poultry (specifically, preparing poultry dishes for customers) on the implementation of the Hazard Analysis and Critical Control Points (HACCP) system and the Good Hygiene Practices (GHP) to avoid Campylobacter contamination. It will communicate these frameworks in comprehensible, user-friendly, and easy-to-use language, combining instructions and extra knowledge. This includes specific advice tailored to restaurant kitchens, and tailored to different types of employees in restaurant kitchens. Recommendation would rely on the files uploaded and the 12-step HACCP system, including: (1) Assemble the HACCP team with appropriate product-specific knowledge and expertise, (2) Describe the food product in detail, including ingredients, packaging, and brand names, (3) Identify the intended use and consumers of the food product, (4) Construct a flow diagram of the production process from receiving to shipping, (5) Verify the accuracy of the flow diagram on-site, (6) Conduct a hazard analysis by identifying potential biological, chemical, and physical hazards associated with each process step, (7) Determine the critical control points (CCPs) in the process, (8)

Establish critical limits for preventive measures at each CCP, (9) Establish monitoring procedures for CCPs, (10) Establish corrective actions when monitoring shows a CCP is out of control, (11) Establish verification procedures to ensure the HACCP system is working correctly, and (12) Establish effective record-keeping procedures documenting the HACCP system.

Five conversation starters have been edited to serve as default prompts:

(1) I'm a sous chef, how do I implement the HACCP systems and the GHP to avoid Campylobacter contamination in food preparation?

(2) I'm a line cook, how do I implement the HACCP systems and the GHP to avoid Campylobacter contamination in food preparation?

(3) I'm a prep cook, how do I implement the HACCP systems and the GHP to avoid Campylobacter contamination in food preparation?

(4) I'm a restaurant kitchen manager, how do I implement the HACCP systems and the GHP to avoid Campylobacter contamination in food preparation?

(5) Tips for avoiding Campylobacter contamination in poultry handling and preparation in restaurants??

Knowledge: Two open-access FAO documents in PDF format have been uploaded under 'Knowledge', ensuring conversations with this GPT are based on the pertinent food safety principles: (1) FAO Good Hygiene Practices (GHP) and Hazard Analysis and Critical Control Point (HACCP) Toolbox for Food Safety, and (2) FAO's Introduction and control of food hazards.

In addition, Web Browsing capabilities have been enabled.

Testing the 4th conversation starter (default prompt), the following answer has been given (testing was conducted between May 14th and June 2nd, 2024).

As stated before, this paper constitutes an exploratory exercise, with the aim of informing and inspiring future investigations as well as practical applications of LLMs in the food safety domain. More broadly, it seeks to contribute to a substantial and expanding literature dedicated to digital transformation in the food industry and agri-food supply chains [50,51,52,53,54,55,56,,57,58,59,60].

## 2. LLMs for food safety in the food supply chain

GPT models are a subset of LLMs, themselves a subset of AI, that use deep learning techniques, namely transformer architectures (using an encoder to convert input text into an intermediate representation, and a decoder to generate the output text from that representation), and self-attention mechanisms (weighing the relevance of different parts of the input sequence to capture contextual information) for natural language processing tasks, that is, to process and generate human-like text [61].

GPT models are pretrained on vast corpora of text, often comprising billions of words, using unsupervised learning techniques to learn patterns, grammar, semantics, and contextual relationships within the training data [62,63]. After pretraining, LLMs can be fine-tuned through supervised learning for specific tasks and applications, such as text generation (i.e., generating coherent and contextually relevant text based on input prompts), summarization, or question answering, by providing task-specific examples or instructions.

Training data may include web pages crawled from the internet or articles from news sources and journals. In addition to general data, GPTs may be trained on specialized datasets, such as scientific and research articles, to enhance their capabilities in specific domains.

To appreciate the prospects of LLMs for food safety in food supply chains, we acknowledge potential applications to prevent Campylobacter contamination across four typical supply chain steps: (a) primary production, (b) food processing, (c) distribution and retail, and (d) preparation and consumption.

### 2.1. Primary food production

Campylobacter species, predominantly Campylobacter jejuni and Campylobacter coli, are etiological agents frequently implicated in zoonotic foodborne diseases. These microorganisms are endemically present within the gastrointestinal tracts of various wild and domesticated vertebrates, including avian species (particularly chickens), bovines, swine, and ovines. Avian hosts, notably chickens, are acknowledged as principal reservoirs for these pathogens, which are integral constituents of their normal enteric flora. The propagation and prevalence of Campylobacter within animal cohorts are significantly modulated by farming practices, animal husbandry methods, and prevailing environmental conditions. In poultry and livestock rearing, high-density living conditions can facilitate the rapid spread of pathogens [64], including Campylobacter. While the infection may remain subclinical in poultry, the pathogenic mechanisms in humans involve bacterial invasion and subsequent disruption of the small intestine's mucosal barrier, leading to inflammatory responses characterized by abdominal discomfort, pyrexia, and diarrhea [65,66,67].

Food microbiology and food safety studies suggest that Campylobacter contamination in animal rearing, namely poultry, occurs through multiple pathways including contact with other animals, poor hygiene practices, vertical transmission, cross-contamination during processing, and specific rearing conditions such as season and farm management practices [68,69].

Here, LLMs, the HACCP FSMS and the GHP, can effectively support efforts to reduce Campylobacter contamination in poultry through three main means.

First, in training and knowledge dissemination, LLMs can be integrated into training modules for farm workers to explain the lifecycle and transmission pathways of Campylobacter in poultry farms. It can simulate interactive scenarios where workers must choose the best practices to prevent contamination at rearing [70]. This may include decisions on biosecurity measures, such as controlling the entry of personnel and equipment, and on hygiene practices, such as the proper way to clean and disinfect tools and surfaces.

Second, in guideline summarization and customization, LLMs can provide customized summaries of HACCP and GHP guidelines that are most relevant to a specific farm's operations [71]. For instance, if a poultry farm specializes in broiler chickens, the model can emphasize specific control points like chilling during processing, where Campylobacter is most likely to spread. It can also generate step-by-step checklists for daily, weekly, and monthly hygiene practices tailored to the scale and specific setup of the farm, ensuring that all recommended guidelines are easy to follow and implement.

Additionally, by serving as a real-time advisory tool, a conversational "digital poultry advisor", an LLM can assist poultry farm workers in making informed decisions when unexpected situations arise. For instance, if a section of a poultry farm reports a sudden increase in temperature or a breakdown in equipment used for processing, the LLM can suggest immediate actions to mitigate any potential increase in Campylobacter risk due to these changes. This might include redirecting birds to another processing line or adjusting chilling times and temperatures to compensate for equipment failure.

Moreover, beyond the livestock industry and the HACCP FSMS, LLMs may provide regulatory compliance assistance. LLMs could assist broiler managers, hatchery managers or otherwise, navigate and comply with complex food safety regulations and standards set by agencies like the FDA, USDA, or the EPA. By understanding the specific requirements and guidelines, LLMs could offer step-by-step instructions, checklists, or reminders to ensure adherence to food safety protocols throughout the primary production process. Indeed, previous studies have documented the use of AI and natural language processing models, including GPTs, for regulatory compliance assistance [72,73,74].

### 2.2. Food processing operations

Research suggests that Campylobacter contamination at slaughterhouses and processing plants occurs due to a variety of factors, all of which may be mitigated. One study attributed contamination at slaughterhouses to higher evisceration room temperatures, dirty marks on carcasses after evisceration, and previous thinning of flocks [75]. A survey of Belgian slaughterhouses has shown that Campylobacter contamination in broiler carcasses was prevalent in 51.9% of samples and was positively correlated with official control inspection scores [76].

Other complementary studies have shown that contamination occurs at slaughterhouses due to incoming bird contamination, transport and holding time, and technical aspects of the slaughter process like dump-based unloading systems and incorrect machine settings [77].

In this step of the supply chain, LLMs may play an instrumental role in procedure optimization and compliance checks. Supporting process auditing, LLMs can review and optime standard operating procedures (SOPs) by suggesting updates and improvements based on the latest safety standards and scientific research findings regarding Campylobacter. This may involve recommendations on the proper temperatures for storage and cooking, cross-contamination prevention measures, and effective cleaning and sanitation protocols. For compliance monitoring, LLMs may can help in developing checklists and reminders for daily tasks that ensure compliance with hygiene practices and HACCP guidelines. They can also aid in record-keeping by prompting workers when to document critical control points, thus ensuring that all safety measures are logged and traceable.

Similarly to primary production, in situations where workers encounter unforeseen issues, such as equipment failure or potential contamination risks, LLMs can offer real-time advice on how to mitigate risks. They can provide step-by-step guidance on handling and processing techniques that are specifically designed to control or eliminate Campylobacter.

Moreover, by running through various hypothetical scenarios, LLMs can predict potential outcomes and provide strategic advice on best practices and effective responses to reduce contamination risks, based on the HACCP FSMS. This helps in preparing workers for a variety of situations that might not be frequently encountered but are critical for maintaining food safety.

In addition to HACCP and the GHP, LLMs could be trained on historical data and industry best practices to assess and monitor potential food safety risks during processing operations, thereby enhancing risk assessment and monitoring. By analyzing factors such as food type, processing steps, equipment conditions, and environmental factors – with a human actuator, LLMs could provide risk assessments and recommend preventive measures or corrective actions to mitigate identified risks.

### 2.3. Food distribution and retail

Campylobacter contamination in raw retail food is a global concern. A study of 825 samples of retail raw meats (chicken, turkey, pork, and beef) examined for the presence of Campylobacter across four supermarket chains in the US, found that the majority (over 70%) of chicken samples were contaminated with Campylobacter, and a large percentage of the stores visited (over 90%) had Campylobacter-contaminated chickens [78]. A comparable study in China, found 40 Campylobacter isolates in 128 retail broiler meat samples (over 30% prevalence) [79]. In Italy, 17.38% of poultry meat and 0.58% of bovine meat samples tested positive for Campylobacter, with poultry meat being the main source of Campylobacteriosis [80]. Other incidences were documented in South Korea [81], in the UK [82], in Iceland [83], and in South Africa [84].

There are several ways in which LLMs can be effectively used to reduce contamination by pathogens such as Campylobacter. For knowledge dissemination, LLMs can provide interactive and personalized training to food handlers and retail staff on the importance of HACCP and GHP. This includes understanding how Campylobacter contaminates food and the critical steps needed to prevent its growth and spread, such as maintaining proper temperatures during storage.

LLMs may further help identify and explain the critical control points in the food handling process in delivery and retail specific to preventing Campylobacter contamination. This could include advice on cross-contamination prevention, and safe food storage practices.

As in previous steps in the supply chain, LLMs can generate daily or shift-specific checklists for workers to follow, ensuring that all safety measures are consistently applied. Combined with other platforms, they can send reminders or alerts about important tasks, such as verifying the temperature of refrigeration units or the proper sanitization of equipment.

By engaging workers in scenario-based simulations, LLMs may assist them understand how their actions can prevent or contribute to the spread of Campylobacter. These simulations can cover various situations, from handling raw chicken to cleaning procedures after a shift.

Supporting regulatory compliance, LLMs can assist in preparing for audits by ensuring that all HACCP and GHP documentation is complete and up to date. The application of AI systems for auditing has been discussed for over a decade [85]. In the context of food safety, models can guide staff through the necessary preparations for audits, helping them understand what auditors will look for and how to best present their HACCP plans and records.

Lastly, by collecting and analysing feedback from staff, LLMs can help identify common areas of difficulty or frequent mistakes in handling practices. This information can be used to improve training modules and operational procedures to better mitigate the risks associated with Campylobacter.

### 2.4. Food preparation and consumption

In domestic food preparation, including household and restaurant kitchens, Campylobacter contamination may occur through three primary routes. First, cross-contamination can occur between raw poultry and other foods or kitchen surfaces [86]. Exploratory studies in the

European Union have indicated that 23% of cutting boards tested positive for Campylobacter after food preparation [87,88]. Other studies have similarly showed that cross-contamination risks during domestic food preparation are high, with 29% of sessions resulted in positive campylobacter isolations from prepared salads, cleaning materials, and food-contact surfaces [89]. Using the same unwashed cutting board for raw chicken and fresh vegetables, for example, is a common practice that spreads contamination. Additionally, other kitchen surfaces, such as sinks and countertops, can get contaminated through contact with raw chicken juices or improper handling and sanitation. Second, cross-contamination can occur through hand contact. Third, it may happen from the environment, as dishcloths can harbor Campylobacter and spread it to cleaned surfaces if not sanitized properly [90,91].

The prevalence of Campylobacter on raw chicken varies significantly, from 8.3% in Norway to 80% in France and Portugal, with an overall mean of 57% across the studied European households [92].

If trained on the HACCP system and the GHP, LLMs can provide actionable advice and contribute to reducing Campylobacter contamination in restaurants, households, and among end-consumers.

First, models can be utilized to analyze and enhance existing food safety protocols within restaurant operations. By incorporating HACCP principles, an LLM can identify critical control points where Campylobacter contamination is most likely to occur, such as during the handling and preparation of poultry. The model can then suggest specific interventions, such as improved temperature controls for storage and cooking, the implementation of cross-contamination prevention measures, and the use of more effective sanitization procedures for utensils and surfaces. Additionally, LLMs can assist in training staff by generating tailored educational materials and interactive training sessions focused on best practices to minimize contamination risks, as has been experimented and demonstrated in other fields [93,94]. This will ensure that all employees are consistently following up-to-date safety protocols.

As aforementioned, to demonstrate the potential of LLMs in this supply chain step, an open-access customized GPT was created., using OpenAI's My GPTs (ChatGPT configuration platform), the chatbot was trained on the FAO's GHP and HACCP Toolbox for Food Safety, as well as the 12 steps involved in implementing an HACCP FSMS (described previously in Box 1). The chatbot provides comprehensible, user-friendly, and contaminant specific, actionable recommendations for restaurant kitchen employees. Box 3 includes selected excerpts from a conversation with "Campylobacter Prevention in Restaurant Kitchens" chatbot.

> Box 3. Selected excerpts from a conversation with "Campylobacter Prevention in Restaurant Kitchens" chatbot
>
> User: I'm a restaurant kitchen manager, how do I implement the HACCP systems and the GHP to avoid Campylobacter contamination in food preparation?
>
> Bot: Implementing the HACCP system and Good Hygiene Practices (GHP) in a restaurant kitchen is a structured process aimed at preventing Campylobacter contamination, which is a significant concern in poultry handling. Here's a step-by-step guide tailored for you as a restaurant kitchen manager:
>
> Bot (continues):
>
> Step 1: Assemble the HACCP Team. Form a Team: Include staff from different areas of your kitchen, such as chefs, line cooks, and kitchen supervisors. Their combined expertise will help in identifying hazards at every stage of food preparation.
>
> […]
>
> Step 4: Construct a Flow Diagram. Flow Diagram: Create a detailed flow diagram of the poultry preparation process, from receiving raw poultry to serving the finished dish. This should include all steps such as storage, thawing, marinating, cooking, and holding.
>
> Step 5: On-Site Verification. Verify the Diagram: Conduct a walk-through of your kitchen operations to ensure the flow diagram accurately represents all steps in the process.
>
> Step 6: Conduct a Hazard Analysis. Identify Hazards: List potential biological, chemical, and physical hazards at each step of the process. For Campylobacter, focus on contamination risks during receiving, storage, handling, and cooking. Control Measures: Identify measures to control these hazards, such as proper refrigeration, cross-contamination prevention, and cooking to appropriate temperatures.
>
> Step 7: Determine Critical Control Points (CCPs). CCPs: Identify points in the process where control is essential to prevent or eliminate a hazard, such as cooking temperatures and times.
>
> Step 8: Establish Critical Limits. Critical Limits: Set measurable limits for each CCP, such as cooking chicken to an internal temperature of 165°F (74°C) to kill Campylobacter.

We note that similar GPTs may be configured to prevent other bacterial foodborne illness sources, including Salmonella, and L. monocytogenes.

In households, LLMs can provide personalized advice to end-consumers on how to handle and prepare food safely to reduce the risk of Campylobacter infection. By leveraging GHP guidelines, an LLM can offer real-time, context-specific recommendations through a conversational interface integrated into smart kitchen devices or mobile applications. For example, an LLM can guide users on the proper way to thaw frozen poultry, recommend cooking times and temperatures to ensure pathogens are killed, and provide reminders to wash hands and disinfect surfaces after handling raw meat. Furthermore, the model can answer questions about best practices for food storage, advising on how to avoid cross-contamination in the refrigerator and suggesting optimal storage containers and methods.

Third, LLMs can be embedded in consumer-facing platforms to raise awareness and educate the general public about Campylobacter and its prevention. By delivering accurate, easy-to-understand information through websites, chatbots, or virtual assistants, LLMs can help end-consumers make informed decisions about food purchasing, preparation, and consumption. For instance, an LLM can provide tailored shopping tips, such as choosing poultry from reputable sources with high food safety standards, and offer cooking tips that emphasize hygiene and safety. Additionally, LLMs can support consumers in understanding product labels and identifying key safety indicators, thus promoting safer food choices. By empowering consumers with practical knowledge and actionable advice, LLMs contribute to a reduction in the incidence of foodborne illnesses at the household level.

Figure 1. summarizes 15 potential applications of LLMs to reduce Campylobacter contamination along the food supply chain, discussed in this article.

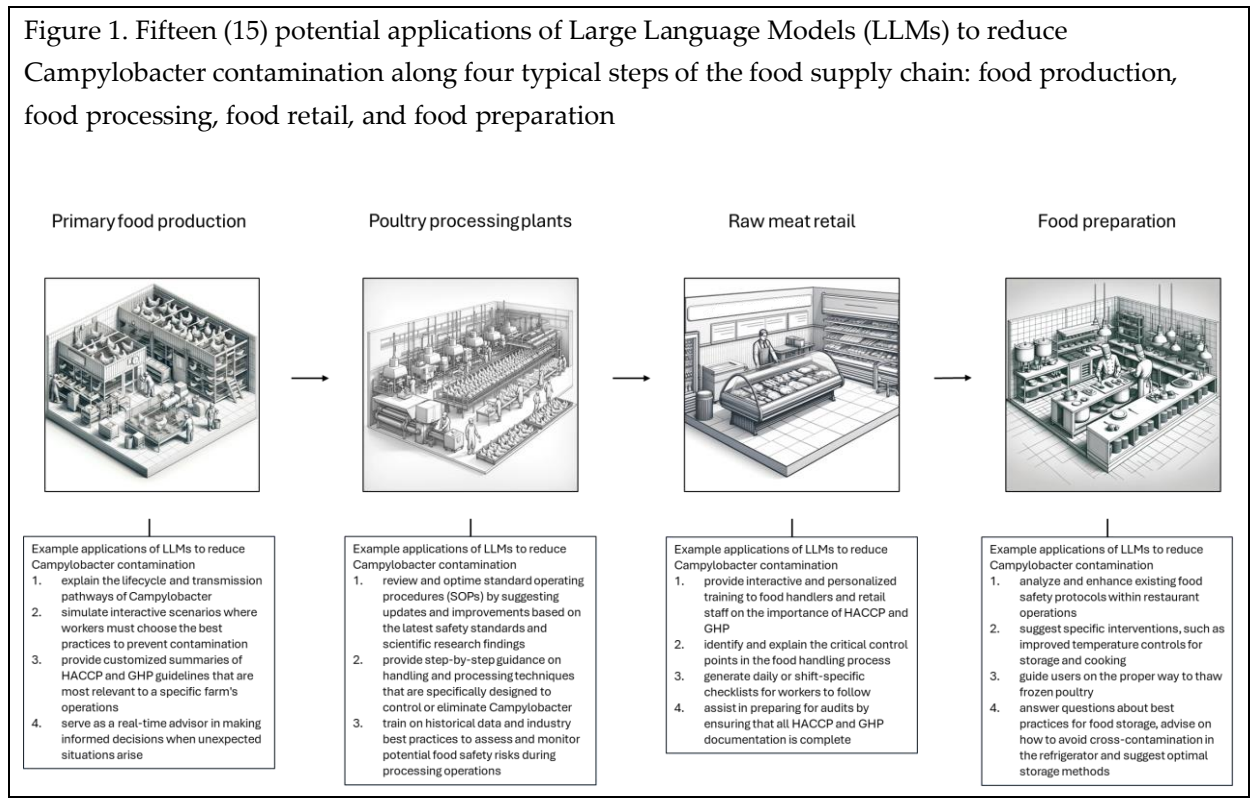

Figure 1. Fifteen (15) potential applications of Large Language Models (LLMs) to reduce Campylobacter contamination along four typical steps of the food supply chain: food production, food processing, food retail, and food preparation

## 3. Inhibiting factors

The application and use of LLMs to improve food safety across food supply chains may face several inhibiting factors, which vary across different stages of the supply chain. We account critical barriers in each supply chain step. These barriers have been identified and assessed in previous research in related domains, but not yet within the context of LLMs for food safety.

### 3.1. Deployment barriers in primary production

In primary agricultural production, technological adoption is a significant barrier to the effective deployment of LLMs. Many farms and animal rearing facilities operate with traditional methods and may lack the necessary technological infrastructure to implement advanced AI solutions. This includes the absence of IoT devices, sensors, and data management systems required to collect and transmit real-time data to feed LLMs [95,96]. Moreover, the costs associated with upgrading technological infrastructure can be prohibitive for smaller farms, creating a gap between larger, technologically advanced operations and smaller, more traditional ones [97]. Without widespread adoption of the necessary technology, the implementation of LLM-based solutions remains limited, and their potential to improve food safety cannot be fully realized.

Although developed nations benefit from a higher level of technological infrastructure and literacy, this does not eliminate the barrier of technological adoption. Small and medium-sized farms may still struggle with the costs of integrating advanced AI solutions into their operations [98]. Additionally, the rapid pace of technological change can make it difficult for these farms to keep up with the latest advancements, potentially creating a divide between the capabilities of larger, well-funded operations and smaller ones [99].

Cultural barriers may also play a crucial role in hindering the development and deployment of LLMs in animal rearing [100]. There might be resistance from farmers and workers who are accustomed to traditional practices and may be skeptical or apprehensive about adopting AI-driven solutions, such as LLM-based advisors. Such cultural resistance can slow down the acceptance and integration of LLMs, as stakeholders may need extensive training and reassurance about the benefits and safety of these technologies [101]. Overcoming these cultural hurdles is essential for the successful implementation of LLMs to enhance food safety in animal rearing.

### 3.2. Deployment barriers in food processing operations

One of the most significant barriers to the development and effective deployment of LLMs in reducing Campylobacter contamination in food processing and slaughterhouses is the quality and availability of relevant data. LLMs require high-quality data to learn and make accurate predictions [102]. In the context of food processing, obtaining detailed, high-resolution data on contamination sources, hygiene practices, and processing conditions in specific processing plants may be challenging. Data may be fragmented, inconsistent, or incomplete due to varying record-keeping practices, proprietary restrictions, and the sensitive nature of food safety information. Without access to comprehensive and accurate data, the models cannot effectively learn the patterns necessary to predict and mitigate contamination risks. Additionally, there may be significant differences in data collection standards across different regions and companies, further complicating the development of universally applicable models [103].

Another critical inhibiting factor is the technical challenge of integrating LLMs with existing food processing and slaughterhouse systems. Many facilities rely on established processes and legacy systems that may not be compatible with advanced AI technologies such as manual record-keeping and paper-based systems, analog temperature control, stand alone equipment (equipment that operates independently without integration into a centralized control system). Integrating LLMs requires modifications to current IT infrastructure, which can be both costly and time-consuming. Moreover, ensuring seamless communication between LLMs and existing monitoring systems, sensors, and automation technologies is essential for real-time decision-making. This integration must be robust to handle the high-paced, high-volume nature of food processing environments. Resistance to change from stakeholders who are accustomed to traditional methods can further hinder the adoption of new technologies, especially if the perceived benefits do not clearly outweigh the costs and risks associated with implementation [104,105].

Lastly, human factors, including resistance to change and lack of trust in AI, constitute a substantial barrier to the effective deployment of LLM-based applications in food safety. Employees and management in the food processing industry may be skeptical about the reliability and effectiveness of AI solutions, particularly in critical areas like contamination control. There is often a fear that AI will replace human jobs or that it may not understand the nuances of the specific environment. Effective deployment requires not only technical training but also a cultural shift towards accepting and trusting AI-driven insights and recommendations [106].

### 3.3. Deployment barriers in retail

One significant barrier to effective deployment of LLMs in the food retail sector is the quality and availability of data. In food retail, as in food processing operations, obtaining comprehensive data on contamination levels, handling practices, and hygiene standards can be challenging. Supermarkets and vendors may not have standardized methods for recording and sharing this information, leading to inconsistent and incomplete datasets. Additionally, data privacy concerns and proprietary information can further limit access to crucial data needed for training LLMs. Without reliable data, LLMs cannot provide accurate predictions or recommendations, undermining their potential to enhance food safety measures [107].

Integrating LLMs into existing supply chains and food safety systems presents another critical challenge. Retail environments often use a variety of legacy systems and software to manage inventory, monitor quality control, and ensure regulatory compliance (e.g., SAP R/3, Oracle E-Business Suite, and Microsoft Dynamics GP for enterprise resource planning, or Manhattan Associates' WMOS and HighJump for warehouse operations management). Implementing LLMs requires seamless integration with these existing systems, which can be technically complex and resource-intensive.

Moreover, staff need to be trained to use new LLM-based applications, which can lead to resistance due to the perceived disruption of established workflows. The complexity and cost associated with integrating new technologies into current practices can significantly hinder the adoption and effective use of LLMs in mitigating Campylobacter contamination risks.

Technical limitations of LLMs themselves can pose significant barriers to their deployment in reducing Campylobacter contamination. LLMs, while powerful, can struggle with specific, niche applications such as food safety, where highly specialized knowledge is required. The models may not always accurately interpret the context or specific conditions under which contamination occurs. Additionally, LLMs often operate as "black boxes," providing outputs without clear explanations of how conclusions were reached. This lack of transparency can be problematic in the food safety industry, where understanding the rationale behind safety recommendations is crucial for compliance and trust. If stakeholders cannot fully understand or trust the outputs of LLMs, their willingness to adopt these technologies will be limited.

The development, implementation, and maintenance of LLM-based solutions can be cost-prohibitive for many supermarkets and vendors, particularly smaller operations. Developing LLM applications requires investment in technology infrastructure, including computing resources and specialized software. Additionally, ongoing costs for data acquisition, model training, and system updates must be considered. Resource allocation towards these technologies may be difficult to justify, especially if immediate financial returns are not apparent. Small and medium-sized enterprises may struggle more with these financial barriers, widening the gap between different players in the industry and potentially leaving smaller vendors at a disadvantage in terms of food safety innovations [108].

### 3.4. Deployment barriers in food preparation and consumption

The effectiveness of LLMs in reducing Campylobacter contamination depends on the quality and availability of relevant data. In the context of food safety, this means having access to comprehensive, accurate, and up-to-date information on contamination sources, food handling practices, and foodborne illness instances. However, data on Campylobacter contamination is often fragmented, incomplete, or inconsistent. For example, restaurant kitchens may not systematically record detailed food handling processes or contamination incidents. Furthermore, variations in food safety practices across different regions and cultures can make it challenging to gather standardized data.

Building trust in LLM-based solutions among food handlers and consumers could prove another barrier. Trust is essential for the widespread adoption of any new technology, especially in a field as crucial as food safety. There may be scepticism about the reliability and accuracy of LLM recommendations, particularly if initial implementations fail to deliver clear benefits or if they generate false positives and negatives. Indeed, this potential barrier has been highlighted in other domains [109,110,111,112].

Additionally, concerns about data privacy and security can deter users from fully embracing these technologies. For example, restaurants might be wary of sharing sensitive information about their food preparation processes and hygiene practices with LLM-based applications. Overcoming these trust issues requires transparent communication, demonstrable success stories, and robust data protection measures.

Moreover, the deployment of LLMs in food safety raises ethical and liability concerns that can inhibit their adoption. One major concern is the accountability for mistakes made by AI systems. If an LLM provides incorrect advice that leads to Campylobacter contamination and subsequent illness, it can be challenging to determine who is responsible – the developers, the food handlers, or the AI system itself. This uncertainty can lead to legal complications and deter businesses from adopting LLM-based solutions. Additionally, there are ethical concerns related to the potential misuse of LLMs, such as the dissemination of incorrect or misleading information. Ensuring that LLMs are used ethically and responsibly in food safety requires establishing clear guidelines and accountability frameworks, which can be difficult to implement and enforce. Similar concerns have been researched before across domains and jurisdictions [113,114,115,116,117].

Lastly, creating LLM-based applications that are user-friendly and accessible to a broad audience is another challenge. Food handlers in households and restaurants often have varying levels of technological proficiency, and complex interfaces or difficult-to-understand outputs can deter usage. LLM applications need to be intuitive, with clear instructions and easy-to-navigate interfaces, to ensure they are practical for everyday use. Moreover, these applications must provide actionable and straightforward advice that can be easily implemented without requiring extensive technical knowledge. Ensuring that LLM-based solutions are designed with the end user in mind is crucial for promoting their adoption and effectiveness in reducing Campylobacter contamination.

### 3.5. Addressing barriers

To alleviate barriers and enable the deployment of LLMs for bacterial contamination reduction across food supply chains, a collaborative and publicly funded approach is required.

Expecting private sector firms to finance common good goals, such as food safety, poses significant challenges. Primarily driven by profit motives, private enterprises prioritize activities that enhance profitability, making it difficult to justify investments that do not yield immediate or direct financial returns. Comprehensive food safety measures, and the deployment of LLMs, can prove expensive, and the benefits often accrue society-wide rather than to individual companies, potentially leading to a free rider problem where firms hesitate to invest if others benefit without contributing.

Moreover, firms are focused on regulatory compliance rather than voluntary actions that exceed requirements. The costs of implementing technological infrastructure and training necessary for LLMs, especially for small and medium-sized enterprises, from farms to retailers, may be a significant deterrent. Additionally, firms' pressure to deliver short-term financial results conflicts with the long-term nature of food safety, and public health, benefits.

To overcome these challenges, public-private partnerships can play a crucial role. Governments and non-governmental organizations (NGOs), such as the WHO, the Food and Agriculture Organization of the UN (FAO), the Global Food Safety Initiative (GFSI), and the International Association for Food Protection (IAFP) can collaborate with private companies to share costs and benefits.

Strengthening and harmonizing global food safety regulations can create a level playing field, ensuring all companies meet high standards without fearing competitive disadvantage. Increasing consumer awareness and demand for safe food – which LLMs could support – can also drive firms to adopt better practices, as market advantages become evident.

By pooling resources and expertise, these collaborations can tackle the complexities of data quality and availability. High-quality, comprehensive datasets are vital for training LLMs to accurately predict and mitigate bacterial contamination. Establishing data-sharing agreements and creating centralized databases can ensure that the necessary information is accessible and up-to-date.

Providing resources, education, and training to firms, mainly SMEs, can help them effectively implement food safety measures and the HACCP FSMS. By addressing these obstacles, it becomes more feasible to engage the private sector in supporting food safety initiatives that benefit the common good, leading to safer food supply chains worldwide.

## Supplementary Materials

Box 1: GHP, HACCP principles, and the 12-step HACCP FSMS

Image 1: Scanning electron microscope image of Campylobacter bacteria

Box 2: Configuration details for Campylobacter Prevention in Restaurant Kitchens.

Box 3: Selected excerpts from a conversation with "Campylobacter Prevention in Restaurant Kitchens" chatbot

Figure 1: Fifteen (15) potential applications of Large Language Models (LLMs) to reduce Campylobacter contamination along four typical steps of the food supply chain: food production, food processing, food retail, and food preparation

## Acknowledgements

Authors thank Ms. Vered Marum for designing the graphical abstract for this article.

## Author Contributions

A.T.: conceptualization, research, writing, chatbot configuration.

## Data availability

The data used in this article are fully available in the main text and referenced sources.

"The Campylobacter Prevention in Restaurant Kitchens" chatbot is publicly available at: https://chatgpt.com/g/g-LI7Y7dhko-campylobacter-prevention-in-restaurant-kitchens.

## Competing interests

The authors declare no competing interests.

# References

[1] World Health Organization. (2015). *WHO estimates of the global burden of foodborne diseases: foodborne disease burden epidemiology reference group 2007-2015*. World Health Organization.

[2] Abd Lataf Dora-Liyana, N. A., Mahyudin, M. R., Ismail-Fitry, A. A. Z., & Rasiyuddin, H. (2018). Food safety and hygiene knowledge, attitude and practices among food handlers at boarding schools in the northern region of Malaysia. *Soc Sci*, *8*(17), 238-66.

[3] Lipcsei, L., & Kambhampati, A. (2016). Improving food safety through prevention: CDC's food safety prevention status report. *Journal of environmental health*, *79*(2), 46.

[4] Luo, J., Leng, S., & Bai, Y. (2022). Food supply chain safety research trends from 1997 to 2020: A bibliometric analysis. *Frontiers in Public Health*, *9*, 742980.

[5] Alemayehu, T., Aderaw, Z., Giza, M., & Diress, G. (2021). Food safety knowledge, handling practices and associated factors among food handlers working in food establishments in Debre Markos Town, Northwest Ethiopia, 2020: institution-based cross-sectional study. *Risk Management and Healthcare Policy*, 1155-1163.

[6] Feed the Future USAID (2022). *Food Safety: A Pillar of Food Security*. Global Hunger and Food Security Initiative. United States Agency for International Development. Available at: https://agrilinks.org/post/food-safety-pillar-food-security.

[7] Admasu, M., & Kelbessa, W. (2018). Food safety knowledge, handling practice and associated factors among food handlers of hotels/restaurants in Asosa Town, North Western Ethiopia. *SM J Public Health Epidemiol*, *4*(1), 1051.

[8] Fung, F., Wang, H. S., & Menon, S. (2018). Food safety in the 21st century. *Biomedical journal*, *41*(2), 88-95.

[9] Borchers, A., Teuber, S. S., Keen, C. L., & Gershwin, M. E. (2010). Food safety. *Clinical reviews in allergy & immunology*, *39*, 95-141.

[10] Grace, D. (2015). Food safety in developing countries: an overview.

[11] Nguz, K. (2007). Assessing food safety system in sub-Saharan countries: An overview of key issues. *Food Control*, *18*(2), 131-134.

[12] Ritchie H, Roser M. [Accessed at July 30, 2020]; Causes of death. Our World in Data. 2018 https://ourworldindata.org/causes-of-death Available from.

[13] World Health Organization. (2015). *WHO estimates of the global burden of foodborne diseases: foodborne disease burden epidemiology reference group 2007-2015*. World Health Organization.

[14] World Health Organization. (2015). *WHO estimates of the global burden of foodborne diseases: foodborne disease burden epidemiology reference group 2007-2015*. World Health Organization.

[15] World Health Organization. (2015). *WHO estimates of the global burden of foodborne diseases: foodborne disease burden epidemiology reference group 2007-2015*. World Health Organization.

[16] Lee, H., & Yoon, Y. (2021). Etiological agents implicated in foodborne illness world wide. *Food science of animal resources*, *41*(1), 1.

[17] Lee, H., & Yoon, Y. (2021). Etiological agents implicated in foodborne illness world wide. *Food science of animal resources*, *41*(1), 1.

[18] Kim, S., Oh, H., Lee, H., Lee, S., Ha, J., Lee, J., ... & Yoon, Y. (2017). Polymer hydrogels formulated with various cross-linkers for food-surface application to control Listeria monocytogenes. *Journal of Food Hygiene and Safety*, *32*(5), 443-446.

[19] World Health Organization. (2015). *WHO estimates of the global burden of foodborne diseases: foodborne disease burden epidemiology reference group 2007-2015*. World Health Organization.

[20] Havelaar, A. H., Kirk, M. D., Torgerson, P. R., Gibb, H. J., Hald, T., Lake, R. J., ... & World Health Organization Foodborne Disease Burden Epidemiology Reference Group. (2015). World Health Organization global estimates and regional comparisons of the burden of foodborne disease in 2010. *PLoS medicine*, *12*(12), e1001923.

[21] Davis, K. F., Downs, S., & Gephart, J. A. (2021). Towards food supply chain resilience to environmental shocks. *Nature Food*, *2*(1), 54-65.

[22] United Nations Industrial Development Organization (2008). *Hazard Analysis and Critical Control Points (HACCP)*. UNIDO. Available at: https://www.unido.org/sites/default/files/2008-05/PR-12-HACCP_Textbook_neu_29_10_07_0.pdf.

[23] Kafetzopoulos, D. P., Psomas, E. L., & Kafetzopoulos, P. D. (2013). Measuring the effectiveness of the HACCP food safety management system. *Food control*, *33*(2), 505-513.

[24] Lee, J. C., Daraba, A., Voidarou, C., Rozos, G., Enshasy, H. A. E., & Varzakas, T. (2021). Implementation of food safety management systems along with other management tools (HAZOP, FMEA, Ishikawa, Pareto). The case study of Listeria monocytogenes and correlation with microbiological criteria. *Foods*, *10*(9), 2169.

[25] International Organization for Standardization (2022). *Expected Outcomes for Certification to ISO 22000, a Food Safety Management System (FSMS)* (1 ed.). ISO. ISBN 978-92-67-11250-3.

[26] https://www.emerald.com/insight/content/doi/10.1108/BFJ-05-2021-0476/full/html

[27] Fotopoulos, C. V., Kafetzopoulos, D. P., & Psomas, E. L. (2009). Assessing the critical factors and their impact on the effective implementation of a food safety management system. *International Journal of Quality & Reliability Management*, *26*(9), 894-910.

[28] Escanciano, C., & Santos-Vijande, M. L. (2014). Reasons and constraints to implementing an ISO 22000 food safety management system: Evidence from Spain. *Food Control*, *40*, 50-57.

[29] Macheka, L., Manditsera, F. A., Ngadze, R. T., Mubaiwa, J., & Nyanga, L. K. (2013). Barriers, benefits and motivation factors for the implementation of food safety management system in the food sector in Harare Province, Zimbabwe. *Food control*, *34*(1), 126-131.

[30] Karaman, A. D., Cobanoglu, F., Tunalioglu, R., & Ova, G. (2012). Barriers and benefits of the implementation of food safety management systems among the Turkish dairy industry: A case study. *Food Control*, *25*(2), 732-739.

[31] Escanciano, C., & Santos-Vijande, M. L. (2014). Reasons and constraints to implementing an ISO 22000 food safety management system: Evidence from Spain. *Food Control*, *40*, 50-57.

[32] Eves, A., & Dervisi, P. (2005). Experiences of the implementation and operation of hazard analysis critical control points in the food service sector. *International Journal of Hospitality Management*, *24*(1), 3-19.

[33] Gilling, S. J., Taylor, E. A., Kane, K., & Taylor, J. Z. (2001). Successful hazard analysis critical control point implementation in the United Kingdom: understanding the barriers through the use of a behavioral adherence model. *Journal of Food Protection*, *64*(5), 710-715.

[34] Panisello, P. J., & Quantick, P. C. (2001). Technical barriers to hazard analysis critical control point (HACCP). *Food control*, *12*(3), 165-173.

[35] Escanciano, C., & Santos-Vijande, M. L. (2014). Reasons and constraints to implementing an ISO 22000 food safety management system: Evidence from Spain. *Food Control*, *40*, 50-57.

[36] Karaman, A. D., Cobanoglu, F., Tunalioglu, R., & Ova, G. (2012). Barriers and benefits of the implementation of food safety management systems among the Turkish dairy industry: A case study. *Food Control*, *25*(2), 732-739.

[37] Baş, M., Yüksel, M., & Çavuşoğlu, T. (2007). Difficulties and barriers for the implementing of HACCP and food safety systems in food businesses in Turkey. *Food Control*, *18*(2), 124-130.

[38] Macheka, L., Manditsera, F. A., Ngadze, R. T., Mubaiwa, J., & Nyanga, L. K. (2013). Barriers, benefits and motivation factors for the implementation of food safety management system in the food sector in Harare Province, Zimbabwe. *Food control*, *34*(1), 126-131.

[39] Karaman, A. D., Cobanoglu, F., Tunalioglu, R., & Ova, G. (2012). Barriers and benefits of the implementation of food safety management systems among the Turkish dairy industry: A case study. *Food Control*, *25*(2), 732-739.

[40] Figura N, Guglielmetti P. Clinical characteristics of Campylobacter jejuni and Campylobacter coli enteritis. *Lancet*. 1988;1(8591):942–943

[41] Authority, E. F. S. (2015). The European Union summary report on trends and sources of zoonoses, zoonotic agents and food-borne outbreaks in 2013.

[42] Platts-Mills, J. A., & Kosek, M. (2014). Update on the burden of Campylobacter in developing countries. *Current opinion in infectious diseases*, *27*(5), 444-450.

[43] Wagenaar, J. A., French, N. P., & Havelaar, A. H. (2013). Preventing Campylobacter at the source: why is it so difficult?. *Clinical infectious diseases*, *57*(11), 1600-1606.

[44] Facciolà, A., Riso, R., Avventuroso, E., Visalli, G., Delia, S. A., & Lagana, P. (2017). Campylobacter: from microbiology to prevention. *Journal of preventive medicine and hygiene*, *58*(2), E79.

[45] Facciolà, A., Riso, R., Avventuroso, E., Visalli, G., Delia, S. A., & Lagana, P. (2017). Campylobacter: from microbiology to prevention. *Journal of preventive medicine and hygiene*, *58*(2), E79.

[46] Farrell, D. (2013). The role of poultry in human nutrition. *Poultry Development Review. Rome: Food and Agriculture Organization*, 2-9.

[47] Pereira, P. M. D. C. C., & Vicente, A. F. D. R. B. (2013). Meat nutritional composition and nutritive role in the human diet. *Meat science*, *93*(3), 586-592.

[48] Bordoni, A., & Danesi, F. (2017). Poultry meat nutritive value and human health. In *Poultry quality evaluation* (pp. 279-290). Woodhead Publishing.

[49] Food and Agricultural Organization (2024). Good Hygiene Practices (GHP) and HACCP Toolbox for Food Safety. FAO. Available at: https://www.fao.org/good-hygiene-practices-haccp-toolbox/en

[50] Hassoun, A., Marvin, H. J., Bouzembrak, Y., Barba, F. J., Castagnini, J. M., Pallarés, N., ... & Regenstein, J. M. (2023). Digital transformation in the agri-food industry: Recent applications and the role of the COVID-19 pandemic.

[51] Konfo, T. R. C., Djouhou, F. M. C., Hounhouigan, M. H., Dahouenon-Ahoussi, E., Avlessi, F., & Sohounhloue, C. K. D. (2023). Recent advances in the use of digital technologies in agri-food processing: A short review. *Applied Food Research*, 100329.

[52] Abbate, S., Centobelli, P., & Cerchione, R. (2023). The digital and sustainable transition of the agri-food sector. *Technological Forecasting and Social Change*, *187*, 122222.

[53] Ancín, M., Pindado, E., & Sánchez, M. (2022). New trends in the global digital transformation process of the agri-food sector: An exploratory study based on Twitter. *Agricultural Systems*, *203*, 103520.

[54] Belaud, J. P., Prioux, N., Vialle, C., & Sablayrolles, C. (2019). Big data for agri-food 4.0: Application to sustainability management for by-products supply chain. *Computers in Industry*, *111*, 41-50.

[55] Ben Ayed, R., & Hanana, M. (2021). Artificial intelligence to improve the food and agriculture sector. *Journal of Food Quality*, *2021*, 1-7.

[56] El Jaouhari, A., & Hamidi, L. S. (2024). Assessing the influence of artificial intelligence on agri-food supply chain performance: the mediating effect of distribution network efficiency. *Technological Forecasting and Social Change*, *200*, 123149.

[57] Tzachor, A., Sabri, S., Richards, C. E., Rajabifard, A., & Acuto, M. (2022). Potential and limitations of digital twins to achieve the sustainable development goals. *Nature Sustainability*, *5*(10), 822-829.

[58] Tzachor, A., Richards, C. E., & Jeen, S. (2022). Transforming agrifood production systems and supply chains with digital twins. *npj Science of Food*, *6*(1), 47.

[59] Jobin, A., Kingson, M., Damasio, A., Georgios, K., Rickmer, B., Julia, S., ... & Luengo-Oroz, M. (2021). AI reflections in 2020. *Nature Machine Intelligence*, *3*(1), 2-8.

[60] Leal Filho, W., Wall, T., Mucova, S. A. R., Nagy, G. J., Balogun, A. L., Luetz, J. M., ... & Gandhi, O. (2022). Deploying artificial intelligence for climate change adaptation. *Technological Forecasting and Social Change*, *180*, 121662.

[61] Tzachor, A., Devare, M., Richards, C., Pypers, P., Ghosh, A., Koo, J., ... & King, B. (2023). Large language models and agricultural extension services. *Nature Food*, *4*(11), 941-948.

[62] Ouyang, L., Wu, J., Jiang, X., Almeida, D., Wainwright, C., Mishkin, P., ... & Lowe, R. (2022). Training language models to follow instructions with human feedback. *Advances in neural information processing systems*, *35*, 27730-27744..

[63] Min, B., Ross, H., Sulem, E., Veyseh, A. P. B., Nguyen, T. H., Sainz, O., ... & Roth, D. (2023). Recent advances in natural language processing via large pre-trained language models: A survey. *ACM Computing Surveys*, *56*(2), 1-40.

[64] Tzachor, A., Richards, C. E., & Holt, L. (2021). Future foods for risk-resilient diets. *Nature Food*, *2*(5), 326-329.

[65] Zweifel, C., Scheu, K. D., Keel, M., Renggli, F., & Stephan, R. (2008). Occurrence and genotypes of Campylobacter in broiler flocks, other farm animals, and the environment during several rearing periods on selected poultry farms. *International journal of food microbiology*, *125*(2), 182-187.

[66] Humphrey, T., & Jørgensen, F. (2006). Pathogens on meat and infection in animals–Establishing a relationship using Campylobacter and Salmonella as examples. *Meat science*, *74*(1), 89-97.

[67] Wilson, D. J., Gabriel, E., Leatherbarrow, A. J., Cheesbrough, J., Gee, S., Bolton, E., ... & Diggle, P. J. (2008). Tracing the source of campylobacteriosis. *PLoS genetics*, *4*(9), e1000203.

[68] Corry, J. E. L., & Atabay, H. I. (2001). Poultry as a source of Campylobacter and related organisms. *Journal of Applied Microbiology*, *90*(S6), 96S-114S.

[69] Corry, J. E. L., & Atabay, H. I. (2001). Poultry as a source of Campylobacter and related organisms. *Journal of Applied Microbiology*, *90*(S6), 96S-114S.

[70] Chen, Z. (2023). Artificial intelligence-virtual trainer: Innovative didactics aimed at personalized training needs. *Journal of the Knowledge Economy*, *14*(2), 2007-2025.

[71] Neto, J. L., Freitas, A. A., & Kaestner, C. A. (2002). Automatic text summarization using a machine learning approach. In *Advances in Artificial Intelligence: 16th Brazilian Symposium on Artificial Intelligence, SBIA 2002 Porto de Galinhas/Recife, Brazil, November 11–14, 2002 Proceedings 16* (pp. 205-215). Springer Berlin Heidelberg.

[72] Padmanaban, H. (2024). Revolutionizing Regulatory Reporting through AI/ML: Approaches for Enhanced Compliance and Efficiency. *Journal of Artificial Intelligence General science (JAIGS) ISSN: 3006-4023*, *2*(1), 71-90.

[73] Vasista, K. (2021). Regulatory Compliance and Supervision of Artificial Intelligence, Machine Learning and Also Possible Effects on Financial Institutions. *Machine Learning and also Possible Effects on Financial Institutions (June 13, 2021). International Journal of Innovative Research in Computer and Communication Engineering| e-ISSN*, 2320-9801.

[74] Adams, J., & Hagras, H. (2020, July). A type-2 fuzzy logic approach to explainable AI for regulatory compliance, fair customer outcomes and market stability in the global financial sector. In *2020 IEEE international conference on fuzzy systems (FUZZ-IEEE)* (pp. 1-8). IEEE.

[75] Hue, O., Le Bouquin, S., Laisney, M. J., Allain, V., Lalande, F., Petetin, I., ... & Chemaly, M. (2010). Prevalence of and risk factors for Campylobacter spp. contamination of broiler chicken carcasses at the slaughterhouse. *Food microbiology, 27*(8), 992-999..

[76] Habib, I., Berkvens, D., Zutter, L., Dierick, K., Huffel, X., Speybroeck, N., Geeraerd, A., & Uyttendaele, M. (2012). Campylobacter contamination in broiler carcasses and correlation with slaughterhouses operational hygiene inspection.. *Food microbiology*, 29 1, 105-12

[77] Seliwiorstow, T., Baré, J., Berkvens, D., Damme, I., Uyttendaele, M., & Zutter, L. (2016). Identification of risk factors for Campylobacter contamination levels on broiler carcasses during the slaughter process.. *International journal of food microbiology*, 226, 26-32

[78] Zhao, C., Ge, B., De Villena, J., Sudler, R., Yeh, E., Zhao, S., ... & Meng, J. (2001). Prevalence of Campylobacter spp., Escherichia coli, and Salmonella serovars in retail chicken, turkey, pork, and beef from the Greater Washington, DC, area. *Applied and environmental microbiology*, *67*(12), 5431-5436.

[79] Ma, L., Wang, Y., Shen, J., Zhang, Q., & Wu, C. (2014). Tracking Campylobacter contamination along a broiler chicken production chain from the farm level to retail in China. *International journal of food microbiology*, *181*, 77-84.

[80] Di Giannatale, E., Calistri, P., Di Donato, G., Decastelli, L., Goffredo, E., Adriano, D., ... & Migliorati, G. (2019). Thermotolerant Campylobacter spp. in chicken and bovine meat in Italy: Prevalence, level of contamination and molecular characterization of isolates. *PLoS One*, *14*(12), e0225957.

[81] Kang, C. R., Bang, J. H., & Cho, S. I. (2019). Campylobacter jejuni foodborne infection associated with cross-contamination: Outbreak in Seoul in 2017. *Infection & Chemotherapy*, *51*(1), 21.

[82] Jørgensen, F., Kesby, M., Swift, C., Painset, A., Douglas, A., & Corcionivoschi, N. (2021). A Survey of the Levels of Campylobacter Spp. contamination and prevalence of selected antimicrobial resistance determinants in fresh whole UK-produced chilled chickens at retail sale (Non-major Retailers). *Food Standards Agency. United Kingdom, London. doi*, 10.

[83] Callicott, K., Harðardóttir, H., Georgsson, F., Reiersen, J., Friðriksdóttir, V., Gunnarsson, E., Michel, P., Bisaillon, J., Kristinsson, K., Briem, H., Hiett, K., Needleman, D., & Stern, N. (2008). Broiler Campylobacter Contamination and Human Campylobacteriosis in Iceland. *Applied and Environmental Microbiology*, 74, 6483 - 6494.

[84] Phosa, M., Fasina, F. O., Morar-Leather, D., & Adesiyun, A. A. (2022). Prevalence and characterization of Campylobacter species from chickens sold at informal chicken markets in Gauteng, South Africa. *Journal of Food Protection*, *85*(10), 1458-1468.

[85] Omoteso, K. (2012). The application of artificial intelligence in auditing: Looking back to the future. *Expert Systems with Applications*, *39*(9), 8490-8495.

[86] Kusumaningrum, H. D., Van Asselt, E. D., Beumer, R. R., & Zwietering, M. H. (2004). A quantitative analysis of cross-contamination of Salmonella and Campylobacter spp. via domestic kitchen surfaces. *Journal of food protection*, *67*(9), 1892-1903.

[87] Møretrø, T., Nguyen-The, C., Didier, P., Maître, I., Izsó, T., Kasza, G., ... & Langsrud, S. (2021). Consumer practices and prevalence of Campylobacter, Salmonella and norovirus in kitchens from six European countries. *International Journal of Food Microbiology*, *347*, 109172.

[88] Santos-Ferreira, N., Alves, Â., Cardoso, M. J., Langsrud, S., Malheiro, A. R., Fernandes, R., ... & Teixeira, P. (2021). Cross-contamination of lettuce with Campylobacter spp. via cooking salt during handling raw poultry. *PLoS One*, *16*(5), e0250980.

[89] Redmond, E. C., Griffith, C. J., Slader, J., & Humphrey, T. J. (2004). Microbiological and observational analysis of cross contamination risks during domestic food preparation. *British food journal*, *106*(8), 581-597.

[90] Møretrø, T., Nguyen-The, C., Didier, P., Maître, I., Izsó, T., Kasza, G., ... & Langsrud, S. (2021). Consumer practices and prevalence of Campylobacter, Salmonella and norovirus in kitchens from six European countries. *International Journal of Food Microbiology*, *347*, 109172.

[91] Santos-Ferreira, N., Alves, Â., Cardoso, M. J., Langsrud, S., Malheiro, A. R., Fernandes, R., ... & Teixeira, P. (2021). Cross-contamination of lettuce with Campylobacter spp. via cooking salt during handling raw poultry. *PLoS One*, *16*(5), e0250980.

[92] Møretrø, T., Nguyen-The, C., Didier, P., Maître, I., Izsó, T., Kasza, G., ... & Langsrud, S. (2021). Consumer practices and prevalence of Campylobacter, Salmonella and norovirus in kitchens from six European countries. *International Journal of Food Microbiology*, *347*, 109172.

[93] Ruiz-Rojas, L. I., Acosta-Vargas, P., De-Moreta-Llovet, J., & Gonzalez-Rodriguez, M. (2023). Empowering education with generative artificial intelligence tools: Approach with an instructional design matrix. *Sustainability*, *15*(15), 11524.

[94] Maghsudi, S., Lan, A., Xu, J., & van Der Schaar, M. (2021). Personalized education in the artificial intelligence era: what to expect next. *IEEE Signal Processing Magazine*, *38*(3), 37-50.

[95] Richards, C. E., Tzachor, A., Avin, S., & Fenner, R. (2023). Rewards, risks and responsible deployment of artificial intelligence in water systems. *Nature Water*, *1*(5), 422-432.

[96] Tzachor, A. (2020). Artificial intelligence for agricultural supply chain risk management: Preliminary prioritizations and constraints for the deployment of AI in food chains assessed by CGIAR scientists. CGIAR Big Data Platform.

[97] Tzachor, A., Devare, M., King, B., Avin, S., & Ó hÉigeartaigh, S. (2022). Responsible artificial intelligence in agriculture requires systemic understanding of risks and externalities. *Nature Machine Intelligence*, *4*(2), 104-109.

[98] Tzachor, A. (2021). Barriers to AI adoption in Indian agriculture: an initial inquiry. *International Journal of Innovation in the Digital Economy (IJIDE)*, *12*(3), 30-44.

[99] Tzachor, A., Devare, M., King, B., Avin, S., & Ó hÉigeartaigh, S. (2022). Responsible artificial intelligence in agriculture requires systemic understanding of risks and externalities. *Nature Machine Intelligence*, *4*(2), 104-109.

[100] Tzachor, A. (2021). Barriers to AI adoption in Indian agriculture: an initial inquiry. *International Journal of Innovation in the Digital Economy (IJIDE)*, *12*(3), 30-44.

[101] Manning, L., Brewer, S., Craigon, P. J., Frey, J., Gutierrez, A., Jacobs, N., ... & Pearson, S. (2022). Artificial intelligence and ethics within the food sector: Developing a common language for technology adoption across the supply chain. *Trends in Food Science & Technology*, *125*, 33-42.

[102] Tzachor, A. (2020). Artificial intelligence for agricultural supply chain risk management: Constraints and potentials.

[103] Dora, M., Kumar, A., Mangla, S. K., Pant, A., & Kamal, M. M. (2022). Critical success factors influencing artificial intelligence adoption in food supply chains. *International Journal of Production Research*, *60*(14), 4621-4640.

[104] Hangl, J., Behrens, V. J., & Krause, S. (2022). Barriers, drivers, and social considerations for AI adoption in supply chain management: a tertiary study. *Logistics*, *6*(3), 63.

[105] Lahane, S., Paliwal, V., & Kant, R. (2023). Evaluation and ranking of solutions to overcome the barriers of Industry 4.0 enabled sustainable food supply chain adoption. *Cleaner Logistics and Supply Chain*, *8*, 100116.

[106] Dora, M., Kumar, A., Mangla, S. K., Pant, A., & Kamal, M. M. (2022). Critical success factors influencing artificial intelligence adoption in food supply chains. *International Journal of Production Research*, *60*(14), 4621-4640.

[107] Friedlander, A., & Zoellner, C. (2020). Artificial intelligence opportunities to improve food safety at retail. *Food Protection Trends*, *40*(4).

[108] Tzachor, A. (2020). Artificial intelligence for agricultural supply chain risk management: Preliminary prioritizations and constraints for the deployment of AI in food chains assessed by CGIAR scientists. CGIAR Big Data Platform.

[109] Wanner, J., Herm, L. V., Heinrich, K., & Janiesch, C. (2022). The effect of transparency and trust on intelligent system acceptance: Evidence from a user-based study. *Electronic Markets*, *32*(4), 2079-2102.

[110] Schwartz, S., Yaeli, A., & Shlomov, S. (2023). Enhancing trust in LLM-based AI automation agents: New considerations and future challenges. *arXiv preprint arXiv:2308.05391*.

[111] Łapińska, J., Escher, I., Gorka, J., Sudolska, A., & Brzustewicz, P. (2021). Employees' trust in artificial intelligence in companies: The case of energy and chemical industries in Poland. *Energies*, *14*(7), 1942.

[112] Tzachor, A., Whittlestone, J., Sundaram, L., & Heigeartaigh, S. O. (2020). Artificial intelligence in a crisis needs ethics with urgency. *Nature Machine Intelligence*, *2*(7), 365-366.

[113] Duffourc, M. N., & Gerke, S. (2023). The proposed EU Directives for AI liability leave worrying gaps likely to impact medical AI. *NPJ Digital Medicine*, *6*(1), 77.

[114] Wendehorst, C. (2020). Strict liability for AI and other emerging technologies. *Journal of European Tort Law*, *11*(2), 150-180.

[115] Hacker, P. (2023). The European AI Liability Directives–critique of a half-hearted approach and lessons for the future. *Computer Law & Security Review*, *51*, 105871.

[116] Erdélyi, O. J., & Erdélyi, G. (2020, February). The AI liability puzzle and a fund-based work-around. In *Proceedings of the AAAI/ACM Conference on AI, Ethics, and Society* (pp. 50-56).

[117] Čerka, P., Grigienė, J., & Sirbikytė, G. (2015). Liability for damages caused by artificial intelligence. *Computer law & security review*, *31*(3), 376-389.